Converse Flexoelectricity of Low-Dimensional Bismuth Selenite (Bi$_2$Se$_3$) Revealed by Piezoresponse Force Microscopy (PFM)

*Qiong Liu,[a] S.S. Nanthakumar,[a] Bin Li,[a] Teresa Cheng,[a] Florian Bittner,[c] Chenxi Ma,[d] Fei Ding,[d] Lei Zheng,[e] Bernhard Roth,[e] and Xiaoying Zhuang[a,b]*

[a]Chair of Computational Science and Simulation Technology, Faculty of Mathematics and Physics, Leibniz University Hannover, Hannover, Germany

[b]Department of Geotechnical Engineering, College of Civil Engineering, Tongji University, Shanghai, China

[c]Institute of Plastics and Circular Economy (IKK), Faculty of Mechanical Engineering, Leibniz University Hannover, Hannover, Germany

[d]Institute of Solid State Physics, Faculty of Mathematics and Physics, Leibniz University Hannover, Hannover, Germany

[e]Hannover Centre for Optical Technologies, Leibniz University Hannover, Hannover, Germany

**Abstract**

Many kinds of two-dimensional (2D) van der Waals (vdW) have been demonstrated to exhibit electromechanical coupling effects, which makes them promising candidates for next-generation devices, such as piezotronics and nanogenerators. Recently, flexoelectricity was found to account for the out-of-plane electromechanical coupling in many 2D transition metal dichalcogenides (TMDs) who only exhibit in-plane piezoelectricity. However, low dimensional vdW three-dimensional (3D) topological insulators (TIs) have been overlooked regarding their electromechanical properties. In this study, for the first time, we experimentally investigate the electromechanical coupling of low dimensional 3D TIs with a centrosymmetric crystal structure, where a binary compound, bismuth selenite ($Bi_2Se_3$), is taken as an example. The results of piezoresponse force microscope (PFM) tests on the $Bi_2Se_3$ nanoflakes show that the material exhibits both out-of-plane and in-plane electromechanical responses. The $Bi_2Se_3$ nanoflake with a thickness of 37 nm possesses an effective out-of-plane piezoelectric coefficient of ~0.65 pm V$^{-1}$. With careful analyses, the electromechanical responses are verified to arise from the converse flexoelectricity. The measured effective out-of-plane piezoelectric coefficient is mainly contributed by flexoelectric coefficient, $\mu_{39}$, which is estimated to be approximately 0.13 nC m$^{-1}$. However, it is rather difficult to obtain the in-plane component of the flexoelectric tensor from the in-plane PFM measurements since the direction of the in-plane stress is always not normal to the AFM cantilever axis. The results provide useful guidance for understanding the flexoelectric effect of low dimensional vdW materials with centrosymmetric crystal structures. Moreover, the work can pave to way to explore the electromechanical devices based on the flexoelectricity of vdW TIs.

**Introduction**

Two-dimensional (2D) van der Waals (vdW) materials are attracting great attention in various applications, owing to their novel mechanical, electrical, magnetic, and optical properties.[1-4] Especially, 2D vdW materials are extremely thin, leading to a considerable flexibility, which makes them ideal platforms to study electromechanical couplings including piezoelectricity and/or flexoelectricity. The electromechanical properties of 2D vdW materials drive them in the arising next-generation devices, such as strain-related sensors and energy harvesters.[5, 6]

A widely used technique to characterize the electromechanical properties of low-dimensional materials is piezoresponse force microscopy (PFM).[7, 8] PFM can detect the electromechanical deformation caused by out-of-plane electric field that is applied by an electrically conducting tip in an atomic force microscopy (AFM). The applied electric field induces a strain, which is referred to as the converse piezoelectric effect. Another type of electromechanical coupling in PFM is converse flexoelectricity, which is the reverse phenomenon of flexoelectricity that describes the appearance of polarization arising from a strain gradient. While the converse flexoelectric effect refers to the appearance of a strain induced by polarization gradient. Unlike piezoelectricity, flexoelectricity or converse flexoelectricity exists in all dielectric materials regardless of their central-symmetry.[7, 8]

Topological insulators (TIs) are a material family that owns metallic conductivity on their boundaries due to the existence of electronic edge (in 2D TIs) or surface states (in three-dimensional (3D) TIs) while exhibits bulk states with energy band gaps.[9,10] Novel phenomena like quantum spin Hall effects and spin momentum locking pave the way for TIs to be as building blocks in quantum computing and spintronics.[9-11] Revealing the electromechanical properties of topological insulators can help explore further applications. For instance, TI piezotronics have been proposed based on the mechanism that piezoelectricity can modulate the electron transport in quantum wells.[12] Investigations of electromechanical properties of 2D vdW materials take much interest in graphene,[13] boron nitride (BN),[14] and transition-metal dichalcogenides (TMDs),[15] and so on, among which graphene is known as a 2D TI.[16] However, less attention was paid to reveal the electromechanical properties, especially the flexoelectricity of other kinds of low-dimensional nanostructures of 3D vdW TIs. Bismuth selenite ($Bi_2Se_3$) is a typical 3D TI with the single Dirac cone, belonging to the V-VI group chalcogenide material family.[9] It is a promising candidate for the realization of spintronic devices.[9] In this study, for the first time, we study the electromechanical properties of thin rhombohedral $Bi_2Se_3$ nanoflakes using the PFM technique. In PFM measurements, it is always difficult to isolate the flexoelectricity from piezoelectricity. The crystal structure of $Bi_2Se_3$ belongs to the ($R$-$3m$) space group, which is centrosymmetric, thus exhibits no bulk piezoelectricity. The flexoelectricity of the $Bi_2Se_3$ nanoflake therefore can be revealed experimentally.

**Results**

The optical images of the as-obtained $Bi_2Se_3$ flakes are shown in Figure 1a. The topography of the thin $Bi_2Se_3$ nanoflakes situated on Au-coated $Si/SiO_2$ substrates were obtained using the contact-mode AFM. Figure 1b is a typical AFM image of a single $Bi_2Se_3$ nanoflake situated in the framed region in Figure 1a. The corresponding line section (Figure 1c) of height shows that its thickness is 37 nm. The investigated $Bi_2Se_3$ has a rhombohedral crystal structure with a centrosymmetric space group ($R$-$3$m). Bulk $Bi_2Se_3$ can be considered to have a layered structure periodically stacked by the so-called quintuple layers with a vdW gap in between every two quintuple layers, as shown in Figure 1d. A single quintuple layer consists of five atomic layers with the sequence -Se2-Bi-Se1-Bi-Se2-. The lattice parameters of $Bi_2Se_3$ are, a = b = 0.413 nm, c = 2.856 nm, α = β = 90°, and γ = 120°. Since each unit cell spans over three quintuple layers, the thickness of a single quintuple is ~0.955 nm.[17] Thus, the $Bi_2Se_3$ nanoflake in Figure 1b is approximately 39 layers.

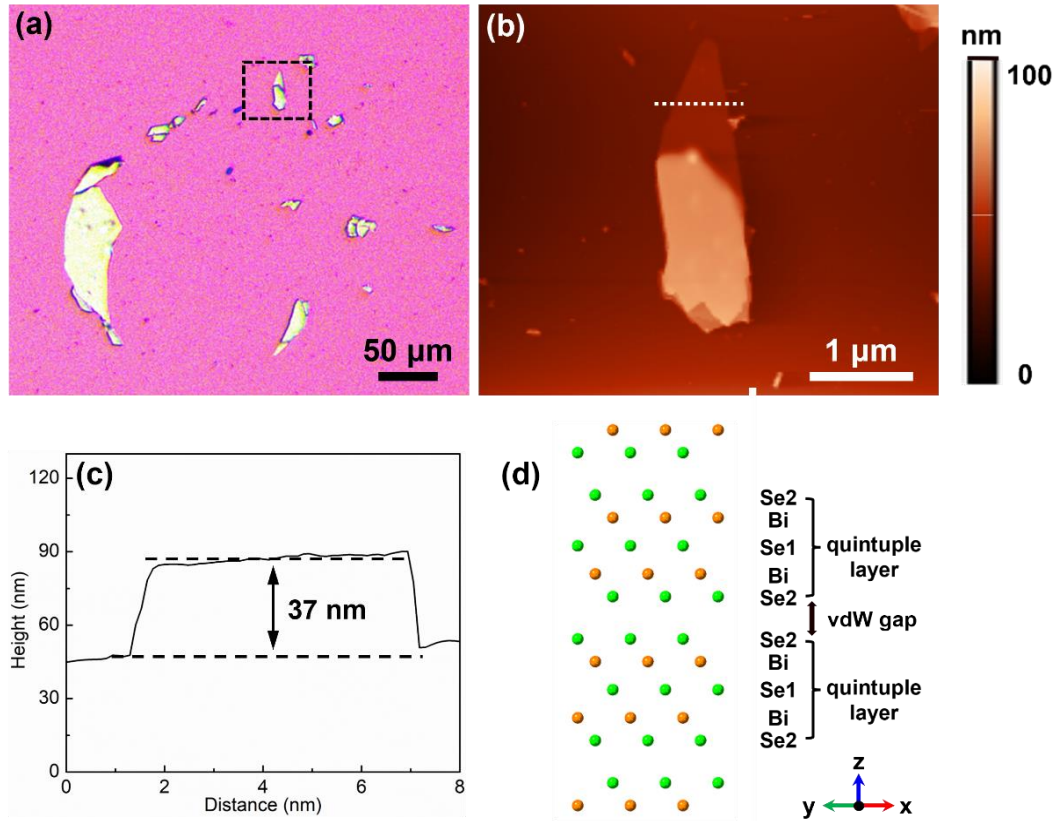

**Figure 1**. Sample characterization. (a) optical image of the as-prepared $Bi_2Se_3$ flakes on an Au-coated $Si/SiO_2$ substrate. (b) contact-mode AFM topography image and (c) the corresponding height profile along the dashed line in panel b of the $Bi_2Se_3$ nanoflake in the framed region in panel a. (d) schematic of the crystal structure of $Bi_2Se_3$.

PFM is often used to characterize piezoelectricity or ferroelectricity by delivering an AC driving voltage ($V_{AC}$) to the surface of the material via an electrically conductive AFM tip and detecting the piezoresponse signal. The experimental setup is schematically shown in Figure 2. The applied AC voltage induces an electric field between the AFM tip and the bottom Au electrode, resulting in the expansion or contraction in the local volume of the material. The displacements of the material cause oscillations of the AFM cantilever, which are detected by a position sensitive photodiode (PSD) and read out with a lock-in amplifier. Since PFM can detect out-of-plane (vertical) and in-plane (lateral) surface displacements, both vertical PFM (VPFM) and lateral PFM (LPFM) techniques are employed to illuminate the electromechanical nature of $Bi_2Se_3$ nanoflakes.

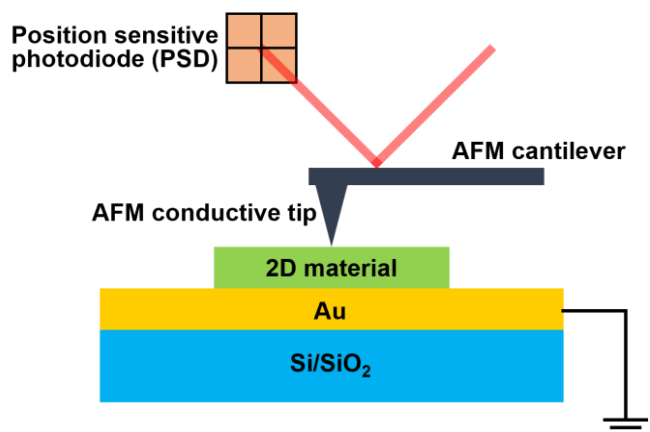

**Figure 2**. A schematic showing the experimental configuration of the PFM tests.

In PFM measurements, the lock-in amplifier can record the amplitude and phase signals while simultaneously acquiring the surface topography. Figure 3 shows the VPFM images under various AC voltages of 0, 3, 5, 7 V. As shown in Figure 3a, the image contrast of 0 V can hardly be observed, indicating that there was no electromechanical response. This suggests that without an external electric field, mechanical displacements cannot be induced. With $V_{AC}$ increasing, the contrast comes out and gets more remarkable, suggesting that the electromechanical coupling arises from the $Bi_2Se_3$ and becomes stronger. Moreover, the electromechanical response signal should not be influenced by crosstalk of the material topographic artifacts since the frequency (60 kHz) of the applied voltage is far from the contact resonance frequency (340 kHz),[15] as shown in Figure S1. Even though there is no contrast in the amplitude image under the AC voltage of 0 V, the topography images acquired during PFM measurements almost remain unchanged (Figures 3a and 3b), which further confirms that the topographic artifacts can be ruled out.

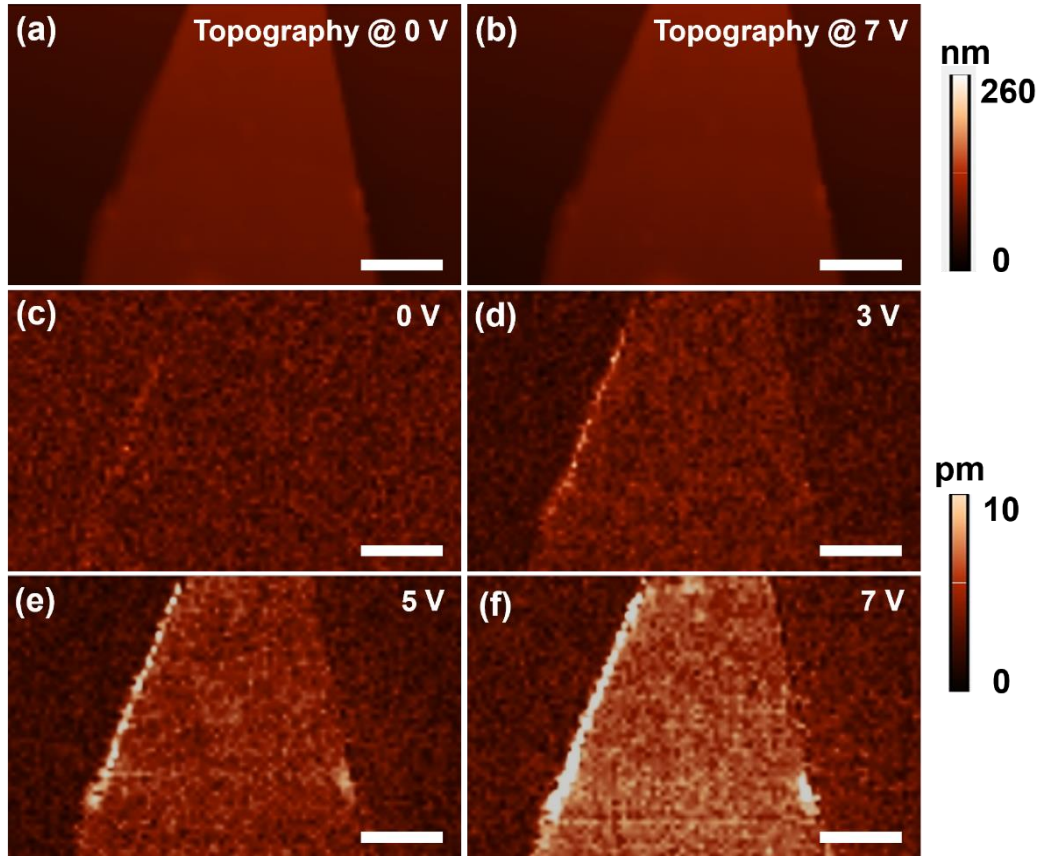

**Figure 3**. VPFM measurements of an individual $Bi_2Se_3$ nanoflake with a thickness of 37 nm. Contact-mode AFM images of the nanoflake under AC voltages of (a) 0 and (b) 7 V, respectively. VPFM amplitude images measured at AC voltages of (c) 0, (d) 3, (e) 5, and (f) 7 V, respectively. Scale bar, 2 μm.

The out-of-plane electromechanical response is usually quantified by the effective piezoelectric coefficient, $d_{33}^{\text{eff}}$, which is given by $d_{33}^{\text{eff}} = A_p/V_{AC}$, where $A_p$ is the deformation displacement. With the deflection sensitivity, $S_d$, of the AFM cantilever measured, $A_p$ can be calculated by converting the electromechanical current signal (AMP) into physical deflection following the equation, $A_p = AMP \times S_d/Gain$.[18] Owing to the internal noise of the device, none-zero amplitude signals always exist and are even detected on the Au substrates with or without driven voltages. The electromechanical response is superimposed by this kind of none-zero background signal.[19] Therefore, a background subtraction method (Figure S2) is used to calculate the actual amplitude arising from the deformation in $Bi_2Se_3$ using both the PFM amplitude and phase channels.[15] Under this method, the PFM image taken on the Au substrate with the driven voltage serves as the background. As can been seen in Figure S3, the phase signals from the substrate and the $Bi_2Se_3$ under each driven voltage show no obvious difference

(<1°). Therefore, the real electromechanical response from $Bi_2Se_3$ can be directly determined by difference between the amplitude values measured on $Bi_2Se_3$ and the substrate.

To more accurately calculate $d_{33}^{\text{eff}}$, the average amplitude value over a specific area for each driven voltage is adopted, which is obtained from the amplitude mapping of this area. Figure 4a shows the relationship between the deflection amplitude and the applied AC voltage. The deflection amplitude can be considered to linearly increase with the applied AC voltage, indicating that $d_{33}^{\text{eff}}$ is independent of the driven voltage. The value of $d_{33}^{\text{eff}}$ is then obtained from the slope of the fitting curve, which is 0.65 ± 0.04 pm V$^{-1}$. This value is close to that (0.7 pm V$^{-1}$) of InSe nanoflakes,[18] but approximately 1/3 smaller than that (1.03 pm V$^{-1}$) of the monolayer $MoS_2$ reported by Yu *et al*,[15] and is much smaller than those of some typical piezoelectric materials, such as 7.5 pm V$^{-1}$ for lithium niobate ($LiNbO_3$) and 3.1 pm V$^{-1}$ for GaN.[15, 20] We performed more PFM tests on $Bi_2Se_3$ flakes with different thicknesses. Each sample shows a linear relationship between the amplitude and the applied AC voltage. However, a thicker $Bi_2Se_3$ flake has a smaller slope, corresponding to a smaller $d_{33}^{\text{eff}}$. Generally, the materials without intrinsic out-of-plane piezoelectricity are most likely to possess a comparative smaller effective out-of-plane piezoelectric coefficient. Thus, the origin of the measured $d_{33}^{\text{eff}}$ will be discussed in the following contents.

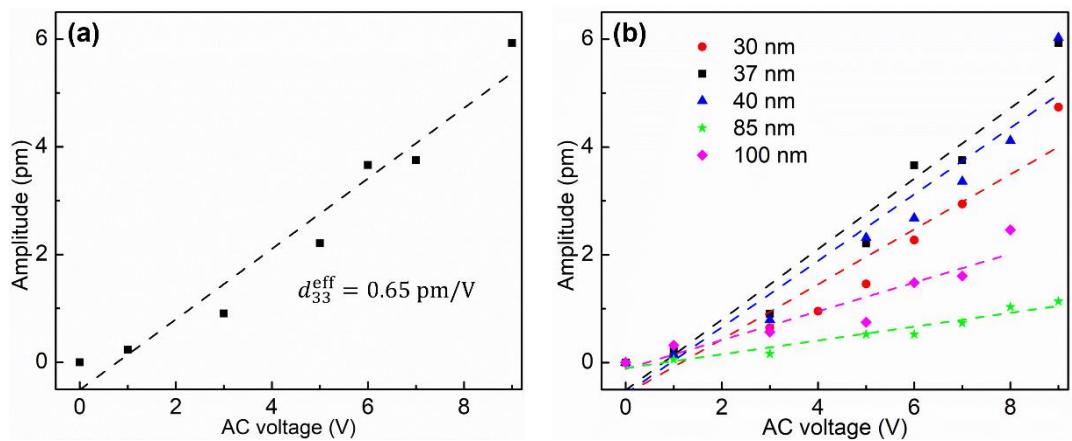

**Figure 4**. Out-of-plane electromechanical properties of $Bi_2Se_3$ nanoflakes. VPFM amplitude as a function of the applied AC voltage for (a) the nanoflake with a thickness of 37 nm and (b) nanoflakes with different thicknesses.

**Discussion**

In PFM measurements, the effective piezoresponse may not be intrinsically contributed by the piezoelectricity of the material since an electric field gradient field is generated below the tip when a voltage is applied. The electric gradient can also generate a mechanical strain, which is called the converse flexoelectric effect.[21] Figure 5 shows the results of the simulated electrostatic field distribution. According to the Coulomb's law, the electric field intensity is intense near the tip and declines with the distance from the tip increasing (Figure 5e), resulting in an inhomogeneous feature (Figure 5c). Simultaneously, the inhomogeneous electric field forms an electric field gradient field, which exhibits a similar pattern as the mother electric field, decaying from near the tip (Figures 5d and 5f). The converse flexoelectricity is another electromechanical coupling mechanism that accounts for the electromechanical response in PFM measurements. Due to the universal existence of flexoelectricity in all materials including those with centrosymmetric crystal structures, a none-zero effective piezoelectric coefficient can be theoretically obtained in any dielectric material.[7] The tested $Bi_2Se_3$ has a centrosymmetric space group of $D_{3d}^5$ (*R-3m*), indicating that the material is none-piezoelectric, and thus there is no out-of-plane piezoelectric coefficients in the material. It is reasonable to assume that the measured $d_{33}^{eff}$ arises from the flexoelectricity. To verify the assumption, careful measures should be taken to exclude other possible contributions to the out-of-plane PFM signals.

Possible effects on the PFM signals from the crosstalk of the surface topography are ruled out as discussed in the contents above. Another possible contribution to the PFM signal amplitude is from the electrostatic force between the tip-cantilever system and the sample, which is described as

$$F_e^{tip} \propto (V_{DC} + V_C)V_{AC} \qquad (1)$$

where $F_e^{tip}$, $V_{DC}$, $V_C$, and $V_{AC}$ represent the electrostatic force, DC tip bias, contact potential difference between the tip and the sample, and AC tip bias, respectively.[22] According to previous works, the interaction between the tip-cantilever and the sample can be suppressed by using a high aspect ratio tip as the one used in out experiments.[22] As shown in Figure 5a, the tip used in this work has a cone ship with a round head whose radius is only ~10 nm (Figure 5b). To confirm that the electrostatic force has a minimal impact on the total PFM signal, we performed the PFM tests on an individual $Bi_2Se_3$ nanoflake by applying different DC biases at a constant AC voltage of 8 V. As shown in Figure S4, the VPFM signal amplitude is independent of $V_{DC}$, indicating that the electrostatic force plays a neglectable role.

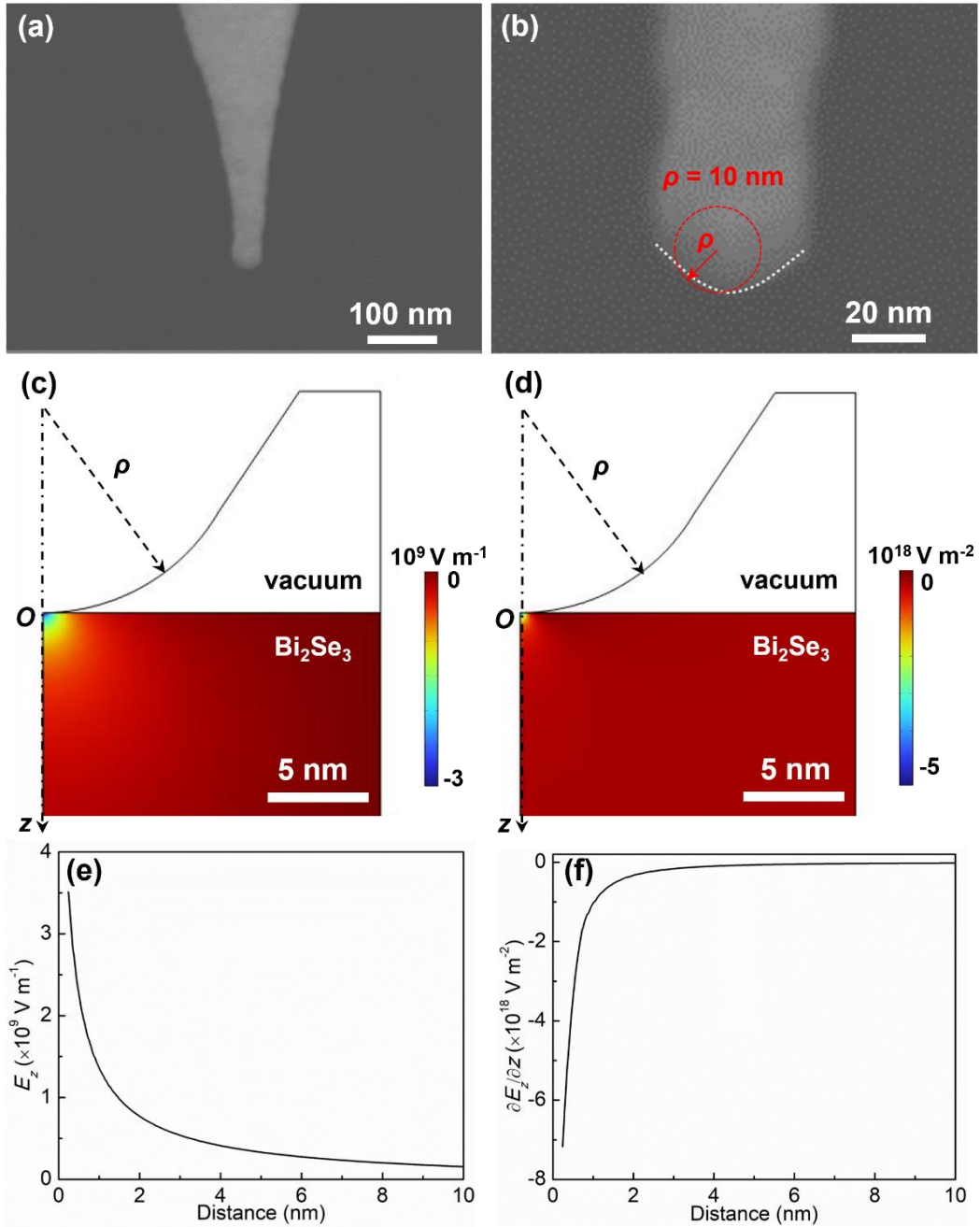

**Figure 5**. Geometry of the conductive AFM tip and contact of the tip with a flat $Bi_2Se_3$. (a) Scanning electron microscopy (SEM) image of the used AFM conductive tip. ρ represents the radius of the tip head, which is 10 nm. (b) zoomed-in SEM image of the framed area in panel a. (c,d) models of distribution of the electric field ($E_z$) and electric field gradient field ($\frac{\partial E_z}{\partial z}$) inside the $Bi_2Se_3$ plate below the tip apex, respectively, under an voltage of 8V. The tip shape is set to be consistent with the one shown in panel b. The thickness of the $Bi_2Se_3$ plate is 10 nm. The dielectric constant of 100 for $Bi_2Se_3$ is adopted according to previous works.[23] (e,f) line curves of the $E_z$ and $\frac{\partial E_z}{\partial z}$ along the z direction.

According to the above analysis, the measured $d_{33}^{eff}$ indeed originates from the converse flexoelectric effect. Similar to the direct flexoelectricity, the converse-flexoelectricity induced stress is also characterized by a fourth rank tensor, which is expressed as

$$\sigma_{ij} = \mu_{ijkl}\frac{\partial E_k}{\partial x_l} \qquad (2)$$

where $\sigma_{ij}$ is the mechanical stress tensor, $\mu_{ijkl}$ is the fourth-order flexoelectric tensor, and $\frac{\partial E_k}{\partial x_l}$ is the electric field gradient tensor. With $ij$ following Voigt notation and $kl$ following 11→1, 12→2, 13→3, 21→4, 22→5, 23→6, 31→7, 32→8, 33→9, the flexoelectric tensor with four indices can be transformed to the tensor with two indices, which is $\mu_{mn}$. Then, the flexoelectric tensor of crystals with a $R$-$3m$ space group is given by[24]

$$\mu_{mn} = \begin{bmatrix} \mu_{11} & 0 & 0 & 0 & \mu_{15} & 2\mu_{63} & 0 & 2\mu_{68} & \mu_{19} \\ \mu_{15} & 0 & 0 & 0 & \mu_{11} & -2\mu_{63} & 0 & -2\mu_{68} & \mu_{19} \\ \mu_{31} & 0 & 0 & 0 & \mu_{31} & 0 & 0 & 0 & \mu_{39} \\ 2\mu_{52} & 0 & 0 & 0 & -2\mu_{52} & \mu_{46} & 0 & \mu_{48} & 0 \\ 0 & \mu_{52} & \mu_{46} & \mu_{52} & 0 & 0 & \mu_{48} & 0 & 0 \\ 0 & \mu_{11}-\mu_{15} & \mu_{63} & \mu_{11}-\mu_{15} & 0 & 0 & \mu_{68} & 0 & 0 \end{bmatrix}$$

(3)

Since the tip is in contact with the surface of the sample and the exciting volume below the tip is pretty small, the electric field inside Bi$_2$Se$_3$ can be considered normal to the sample surface.[15, 25] Then the first six rows of the electric field gradient tensor can be neglected. Correspondingly in Equation 3, the first six columns of the flexoelectric tensor are eliminated. Equation 2 will be simplified as

$$\begin{bmatrix} 0 & 2\mu_{68} & \mu_{19} \\ 0 & 2\mu_{68} & \mu_{19} \\ 0 & 0 & \mu_{39} \\ 0 & \mu_{48} & 0 \\ \mu_{48} & 0 & 0 \\ \mu_{68} & 0 & 0 \end{bmatrix} \begin{bmatrix} \frac{\partial E_3}{\partial x_1} \\ \frac{\partial E_3}{\partial x_2} \\ \frac{\partial E_3}{\partial x_3} \end{bmatrix} = \begin{bmatrix} \sigma_1 \\ \sigma_2 \\ \sigma_3 \\ \sigma_4 \\ \sigma_5 \\ \sigma_6 \end{bmatrix} \qquad (4)$$

Equation 4 indicates that six stress components are deduced, which are

$$\begin{cases} \sigma_1 = 2\mu_{68}\frac{\partial E_3}{\partial x_2} + \mu_{19}\frac{\partial E_3}{\partial x_3} \\ \sigma_2 = 2\mu_{68}\frac{\partial E_3}{\partial x_2} + \mu_{19}\frac{\partial E_3}{\partial x_3} \\ \sigma_3 = \mu_{39}\frac{\partial E_3}{\partial x_3} \\ \sigma_4 = \mu_{48}\frac{\partial E_3}{\partial x_2} \\ \sigma_5 = \mu_{48}\frac{\partial E_3}{\partial x_1} \\ \sigma_6 = \mu_{68}\frac{\partial E_3}{\partial x_1} \end{cases} \quad (5)$$

Among the six stress components in Equation 5, $\sigma_6$ is the in-plane shearing stress, which may lead to buckling of the AFM cantilever, contributing to the out-of-plane PFM response when $\sigma_6$ has a component that is parallel to the AFM cantilever and pointing from the AFM tip to the cantilever end. Nevertheless, a possible induced bucking deformation by $\sigma_6$ should be neglectable due to the insignificant value of $\frac{\partial E_3}{\partial x_1}$ compared to $\frac{\partial E_3}{\partial x_1}$. For instance, as shown in Figure S5, the VPFM signal amplitude from the same area of Bi$_2$Se$_3$ remains nearly the same before and after rotating the sample by 180°, suggesting that there are ignorable contributions from buckling to the vertical PFM signal. Instead, $\sigma_6$ plays a major role in the in-plane PFM signals. $\sigma_1$ and $\sigma_2$ are in-plane stresses, leading to in-plane contraction or expansion of the material. They may contribute to the out-of-plane surface displacements due to Possion-like effects, which, nevertheless, are most likely to make a neglectable contribution. $\sigma_3$, $\sigma_4$, and $\sigma_5$ can directly contribute to the out-of-plane PFM signal amplitude through inducing a deflection of the cantilever. Explicitly, $\sigma_3$ is the out-of-plane stress, which is the product of $\frac{\partial E_3}{\partial x_3}$ and the vertical (out-of-plane) flexoelectric coefficient, $\mu_{39}$; $\sigma_4$ and $\sigma_5$ are out-of-plane shearing stresses, which are generated by $\frac{\partial E_3}{\partial x_1}$ and $\frac{\partial E_3}{\partial x_2}$, respectively, and related to the lateral (in-plane) flexoelectric coefficient, $\mu_{48}$. Compared to $\frac{\partial E_3}{\partial x_3}$, $\frac{\partial E_3}{\partial x_1}$ and $\frac{\partial E_3}{\partial x_2}$ can be assumed to be neglectable. Then the measured $d_{33}^{\text{eff}}$ is considered to be mainly contributed by $\sigma_3$, from which we are able to estimate $\mu_{39}$ according to Equations 5-7,

$$\sigma_3 = C_{33}\varepsilon_3 \quad (6)$$

$$\varepsilon_3 = d_{33}^{\text{eff}} E_3 \quad (7)$$

where $\varepsilon_3$ is the out-of-plane strain, and $C_{33}$ is the elastic constant. Then, $\mu_{39}$ is calculated by

$$\mu_{39} = \frac{C_{33} d_{33}^{\text{eff}} E_3}{E_{3,3}} \quad (8)$$

where $\overline{E_3}$ and $\overline{E_{3,3}}$ are the mean values of $E_3$ ($E_z$) and $\frac{\partial E_z}{\partial x_3}$ ($\frac{\partial E_z}{\partial z}$) through the thickness ($h$) of the sample, and are obtained according to Equations 9 and 10, respectively,

$$\overline{E_3} = \frac{1}{h}\int_0^h |E_z| dz \qquad (9)$$

$$\overline{E_{3,3}} = \frac{1}{h}\int_0^h \left|\frac{\partial E_z}{\partial z}\right| dz. \qquad (10)$$

By adopting a value of ~102 GPa[26] of $C_{33}$ for Bi$_2$Se$_3$ obtained by density-functional theory (DFT) reported previously and getting the ratio of $\overline{E_3}$ to $\overline{E_{3,3}}$, we can calculate $\mu_{39}$ according to Equation 8, which is 0.13 ± 0.01 nC m$^{-1}$. This value is of the same order of magnitude as the measured effective ou-of-plane flexoelectric coefficients of some 2D TMD materials without out-of-plane piezoelectricity. For instance, Shimada et al measured the flexoelectric coefficient $\mu_{39}$ of MoS$_2$ nanoflakes by PFM measurements, which is ~0.27 nC m$^{-1}$.[25] In Yu's works, the effective flexoelectric coefficients that contribute to the out-of-plane electromechanical response of monolayer MoS$_2$, MoSe$_2$ and WS$_2$ are estimated to be ~0.1, 0.065, 0.103, and 0.053 nC m$^{-1}$, respectively.[27] Generally, a material with a larger dielectric constant probably has a stronger electromechanical response. Bi$_2$Se$_3$ has a comparative higher dielectric constant (~100) than the typical investigated TMDs (<10), which is likely to lead to a much larger flexoelectric coefficient.[18, 28] The similar values of the effective out-of-plane flexoelectric coefficients of Bi$_2$Se$_3$ and TMDs are probably due to the larger thicknesses of the Bi$_2$Se$_3$ nanoflakes tested in this work. Previous works reported that the flexoelectric effect increases with the thickness decreasing.[18] The results of the PFM amplitude with different thicknesses in Figure 4b show that thicker Bi$_2$Se$_3$ flakes have smaller $d_{33}^{\text{eff}}$, corresponding to smaller effective $\mu_{39}$, which indicates that the flexoelectricity of 2D Bi$_2$Se$_3$ is thickness dependent. Thus, it can be speculated that with the thickness of Bi$_2$Se$_3$ continuing to decrease, larger effective $\mu_{39}$ is likely obtained. However, the speculation needs further clarification.

As mentioned above, the stress component $\sigma_6$ in Equation 5 will cause in plane shearing deformation. It will contribute to in-plane PFM signals when it has a component that is normal to the AFM cantilever axis. We used LPFM to study the in-plane electromechanical coupling of Bi$_2$Se$_3$ nanoflakes. As shown in Figures 6c-h, the enhanced contrast of the in-plane PFM amplitude images of the sample indicates that the amplitude increases with the AC voltage increasing. Similar to the out-of-plane PFM amplitude signal, there is also a linear relationship between the in-plane amplitude and the AC voltage, as shown in Figure S6. According to Equation 5, the in-plane PFM signal is contributed by the component $\mu_{68}$ of the flexoelectric tensor. However, it is rather difficult to calculate the value of $\mu_{68}$ unless the direction of $\sigma_6$ is normal to the AFM cantilever axis.

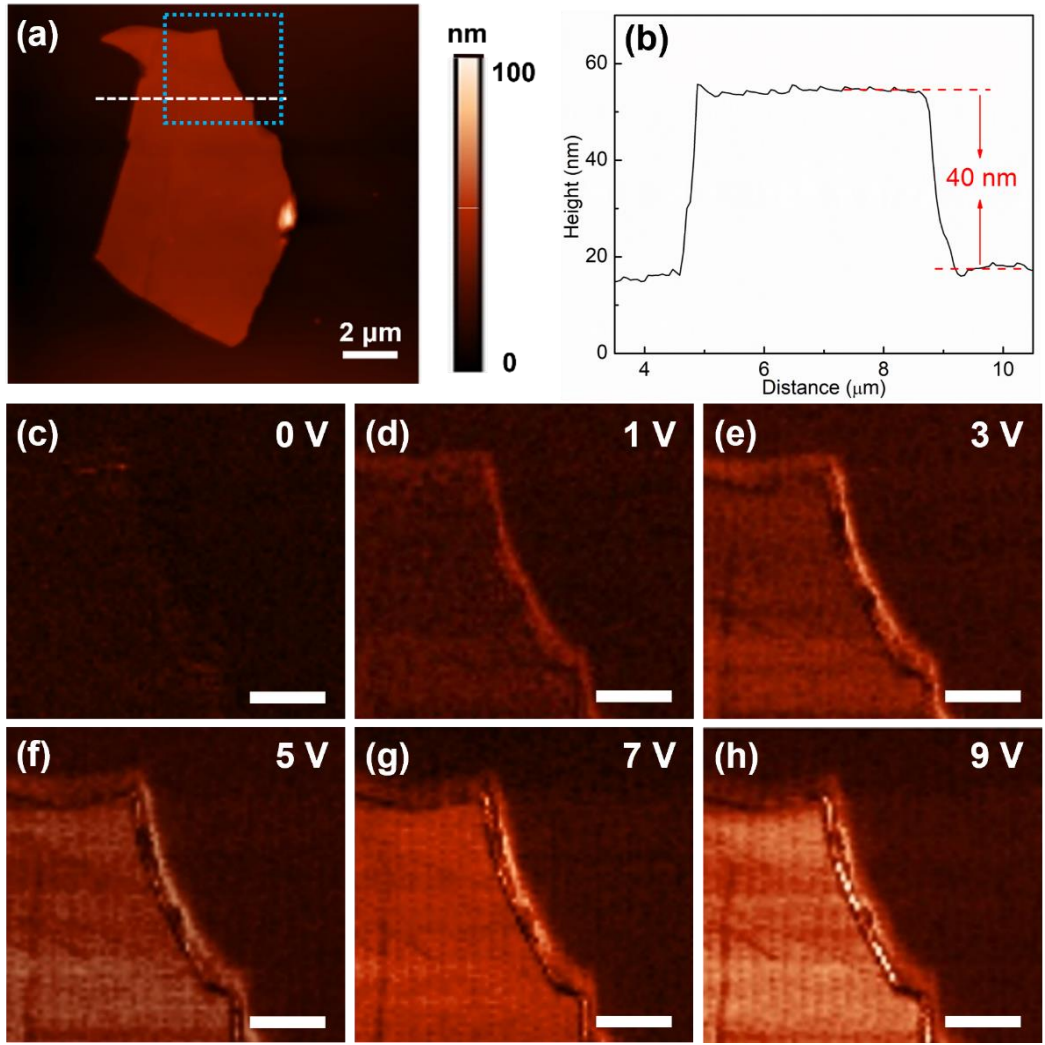

**Figure 6**. LPFM measurements of an individual Bi$_2$Se$_3$ nanoflake with a thickness of ~40 nm. (a) AFM topography of the Bi$_2$Se$_3$ nanoflake and (b) the corresponding line profile of the height along the dashed line in panel a. (c-h) in-plane PFM amplitude images of the framed region in panel a with AC applied voltages of 0, 1, 3, 5, 7 and 9 V, respectively. Scale bar, 1 μm.

**Conclusion**

In summary, for the first time, we shed light on the electromechanical properties of low-dimensional rhombohedral Bi$_2$Se$_3$, which is a 3D vdW TI. The Bi$_2$Se$_3$ nanoflakes with a thickness of 37 nm has an effective out-of-plane piezoelectric coefficient of ~0.65 pm V$^{-1}$. The rhombohedral Bi$_2$Se$_3$ owns a centrosymmetric crystal structure, which shows no piezoelectricity. The out-of-plane and in-plane electromechanical responses are verified to originate from the converse flexoelectricity after careful analyses. The flexoelectric coefficient mainly accounting for the measured out-of-plane effective piezoelectric coefficient is estimated to be ~0.13 nC m$^{-1}$, which is of the same order of magnitude as the measured effective flexoelectric coefficients of some 2D TMD materials without out-of-plane piezoelectricity. However, it is rather difficult to obtain the in-plane component of the flexoelectric tensor from the in-plane PFM measurements since the direction of the in-plane stress is always not normal to the AFM cantilever axis. The results provide useful guidance for understanding the flexoelectric effect of low dimensional vdW materials with centrosymmetric crystal structures. This work may pave the way to apply low-dimensional TIs to novel devices, such as piezotronics and sensors.

**Methods**

Sample fabrication: High-quality (purity > 99.995%) rhombohedral $Bi_2Se_3$ bulk crystals were purchased from HQ Graphene Inc., Netherlands. Low-dimensional $Bi_2Se_3$ flakes were prepared using a facile mechanical exfoliation method,[29] which involves using scotch tapes to exfoliate a bulk $Bi_2Se_3$ crystal repeatedly. Briefly, a bulk $Bi_2Se_3$ crystal was placed on a piece of scotch tape, then another piece of tape was pressed against the crystal and peeled off. The process was repeated for several times. Next, the tape with $Bi_2Se_3$ nanoflakes was pressed against a homemade polydimethylsiloxane (PDMS) stamp that was prepared on a polished Si substrate by spin-coating. After the scotch tape was slowly peeled off, some thin $Bi_2Se_3$ flakes were left on the PDMS stamp. The PDMS stamp with $Bi_2Se_3$ nanoflakes was put against an Au-coated $Si/SiO_2$ substrate, gently pressed, and released. Finally, low-dimensional $Bi_2Se_3$ flakes were left on the Au-coated $Si/SiO_2$ substrate.

Characterization: The geometry of the AFM tip was characterized by a SEM (PIONEER Two, RAITH, Germany).

PFM Tests: PFM measurements were performed in the contact mode on a NTEGRA Prima scanning probe microscope (NT-MDT Co., Ireland) in ambient using commercial conductive Cr/Au coated silicon probes with a spring constant of ~2.8 N m$^{-1}$ and a free resonance frequency of ~75 kHz. AC driven voltages with frequency of 60 kHz were applied between the probe tip and the Au layer on the $Si/SiO_2$ substrate for both vertical PFM (VPFM) and lateral PFM (LPFM) measurements, where the PFM response were measured by the lock-in amplification technique.

Supporting Information

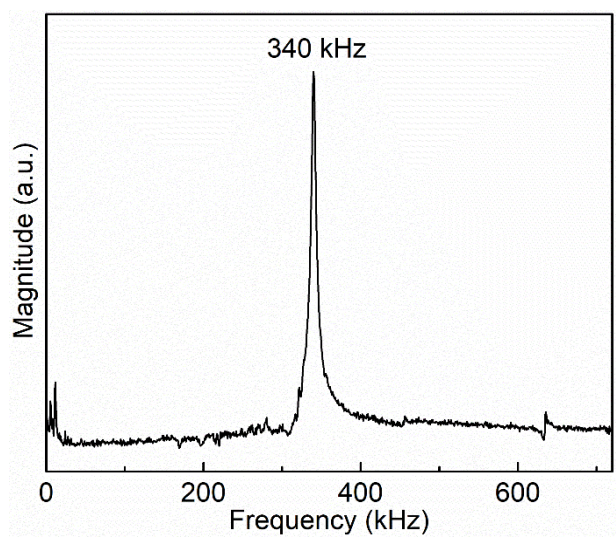

**Figure S1**. Determination of the frequency of the AC voltage for the PFM measurements: the piezoresponse amplitude versus AC voltage frequency measured on the $Bi_2Se_3$ nanoflake situated on the Au-coated $Si/SiO_2$ substrate. In the frequency regime below 30 kHz and between 150 to 300 kHz, the amplitude signal shows a fluctuation. The contact resonance frequency of the AFM cantilever is 340 kHz. In the PFM measurements of the samples, the frequency of 60 kHz for the AC voltage is chosen, which is far from the contact resonance frequency and within the regime for steady amplitude signal.

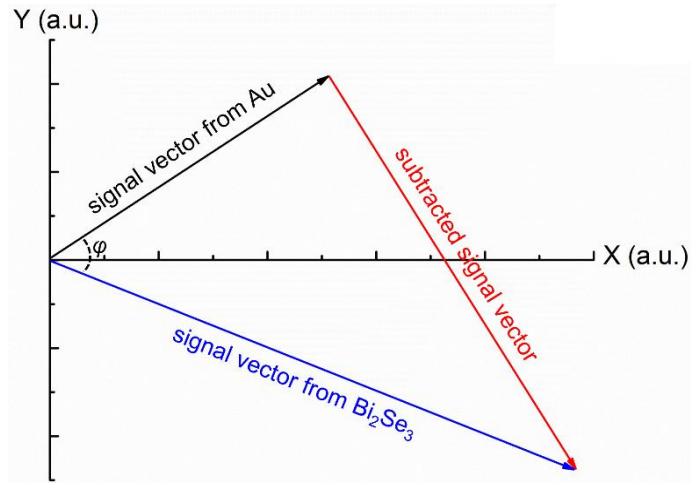

**Figure S2**. A schematic showing the background subtraction method for obtaining the actual PFM amplitude signal of samples. X and Y represent the *x*- and *y*-components of the amplitude, respectively. With the amplitude and phase signals acquired, a vector can be drawn. The black and blue arrows are signal vectors from Au and $Bi_2Se_3$, respectively, with an angle, $\varphi$, in between; by using the blue arrow to subtract the black arrow, a background subtracted signal vector can be obtained, whose magnitude is the actual amplitude originating from $Bi_2Se_3$ and can be calculated according to the law of cosines. When the phase signals from the Au and the sample are close, $\varphi$ is pretty small. In this case, the actual amplitude signal from $Bi_2Se_3$ can be considered to be the difference of the magnitude of the signal vector from $Bi_2Se_3$ and the signal vector from Au.

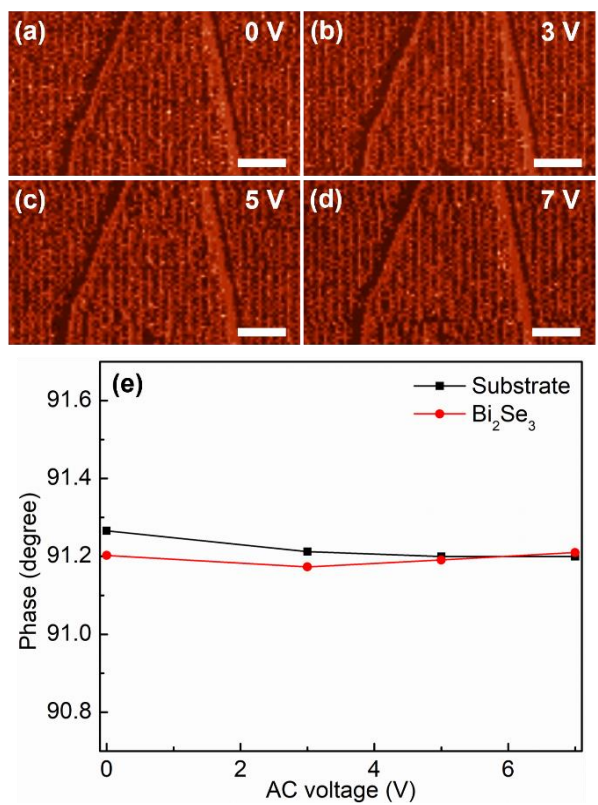

**Figure S3**. VPFM phase results of the Bi$_2$Se$_3$ nanoflake in Figure 3. (a-d) PFM phase images of the sample, and (e) comparison of the phase values from Bi$_2$Se$_3$ and Au measured at AC voltages of (c) 0, (d) 3, (e) 5, and (f) 7 V, respectively. Scale bar, 2 μm.

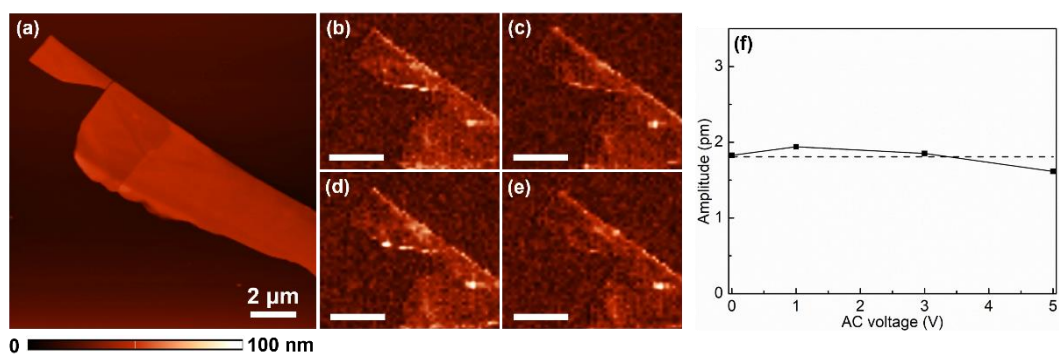

**Figure S4**. VPFM measurements on an individual Bi$_2$Se$_3$ nanoflake with different DC voltage at a constant AC voltage of 8V. (a) AFM topography, (b-e) PFM amplitude images of the nanoflake with DC voltages of 0, 1, 3, and 5 V, respectively. Scale bar, 2 μm. (f) values of PFM amplitude. The dash line represents the mean value.

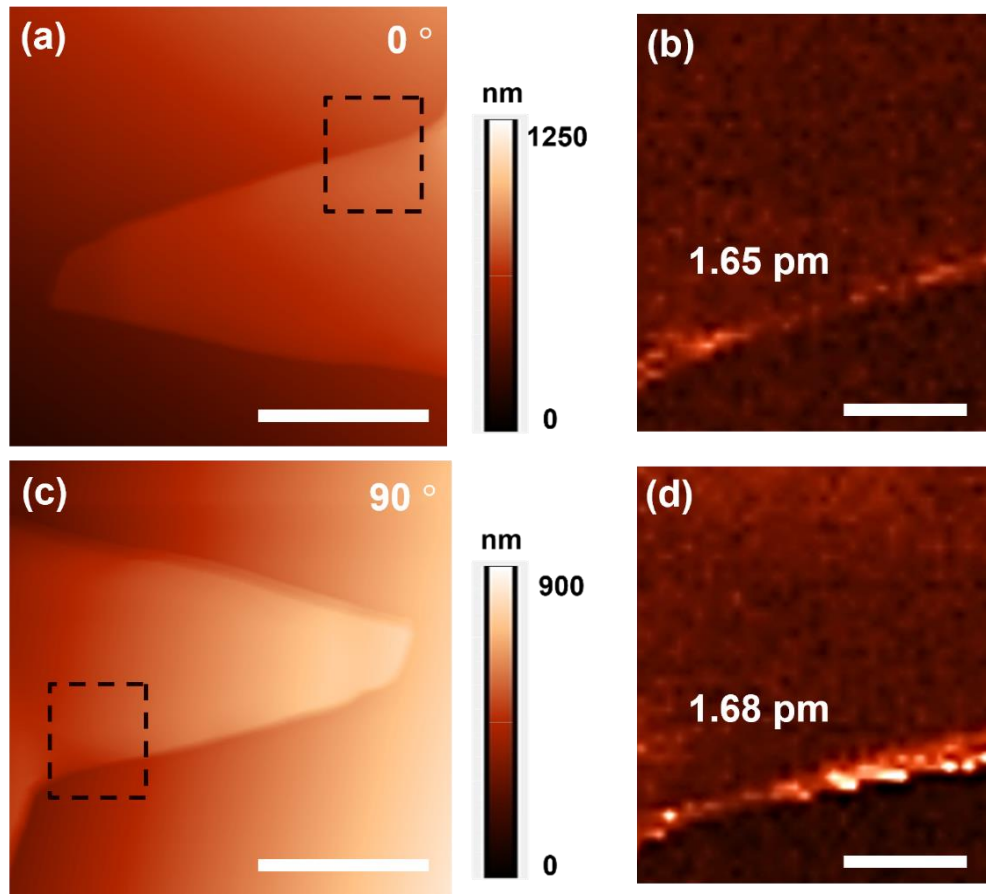

**Figure S5**. VPFM measurements of a Bi$_2$Se$_3$ nanoflake for ruling out the AFM cantilever buckling. (a,c) AFM topography images, and (b,d) PFM amplitude images of the nanoflake before and after rotating it by 180°. The deformation amplitudes are 1.65 and 1.68 pm, respectively. Scale bar, 5 μm in a and c, 1 μm in b and d.

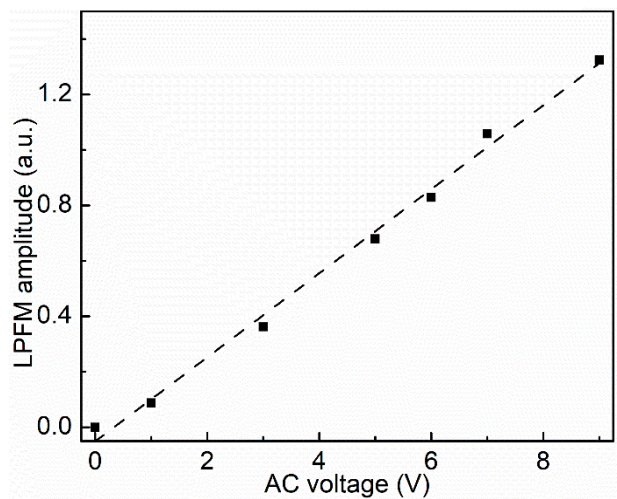

**Figure S6**. Relationship between the LPFM amplitude signal and the applied AC voltage.